\documentclass[
]{ceurart}

\usepackage{listings}
\usepackage{tabularx}
\usepackage{booktabs}
\usepackage{float}
\usepackage{threeparttable}
\usepackage{ragged2e}
\begin{document}

\newcommand{\pmu}[1]{\textcolor{teal}{{pmu: #1}}}
\newcommand{\afo}[1]{\textcolor{orange}{{afo: #1}}}
\newcommand{\dko}[1]{\textcolor{blue}{{afo: #1}}}

\copyrightyear{2026}
\copyrightclause{Copyright for this paper by its authors.
  Use permitted under Creative Commons License Attribution 4.0
  International (CC BY 4.0).}

\conference{Preprint accepted for the Joint Proceedings of the ACM UMAP Workshops 2026, UMAP 2026, June  8–11, 2026, Gothenburg, Sweden}

\title{Fair Agents: Balancing Multistakeholder Alignment in Multi-Agent Personalization Systems}


\author[1]{Andrea Forster}[%
orcid=0009-0008-6818-1916,
email=aforster@know-center.at
]
\cormark[1]
\address[1]{Fair-AI, Know Center Research GmbH, Graz, Austria}
\address[2]{Institute of Human-Centred Computing, Graz University of Technology, Graz, Austria}
\address[3]{Department of Digital Humanities, University of Graz, Graz, Austria}

\author[1]{Peter Müllner}[%
orcid=0000-0001-6581-1945,
email=pmuellner@know-center.at
]

\author[2]{Denis Helic}[%
orcid=0000-0003-0725-7450,
email=dhelic@tugraz.at
]

\author[2]{Elisabeth Lex}[%
orcid=0000-0001-5293-2967,
email=elisabeth.lex@tugraz.at
]

\author[1,3]{Dominik Kowald}[%
orcid=0000-0003-3230-6234,
email=dkowald@know-center.at
]
\cormark[1]

\cortext[1]{Corresponding author.}

\begin{abstract}
LLM agents are increasingly used for personalization due to their ability to communicate directly with users in natural language, integrate external knowledge bases, and negotiate with other (possibly human) agents. Especially in multistakeholder AI systems with multiple distinct objectives, LLM agents are used to independently optimize for each stakeholder's goals. Here, stakeholder alignment is essential to identify and map these goals to provide LLM agents with quantifiable objectives. Plus, the way in which the outputs of the LLM agents are aggregated is fundamental to ensuring fair outcomes for all agents and, therefore, stakeholders.
In this work, we identify open research challenges and propose a conceptual framework for designing fair multi-agent multistakeholder personalization systems that balance competing stakeholder objectives. 
Our framework integrates (i) methods to align stakeholder objectives and LLM agents, (ii) aggregation strategies, e.g., based on social choice theory, to form fair collective decisions, and (iii) stakeholder-centric evaluation procedures for both individual and collective agent behavior. We showcase our framework through a tourism use case and discuss possible applications in other domains, such as education and healthcare. Finally, we discuss domain-specific fairness tensions and review datasets for evaluating multistakeholder fairness and multi-agent personalization systems.
\end{abstract}


\maketitle
\section{Introduction}
Large language model (LLM) agents are reshaping personalized recommender systems~\cite{huang2025towards}. Traditional systems infer preferences from behavioral patterns and rank pre-defined item sets. LLM agents, in contrast, capture user intent through natural language, integrate multimodal inputs, leverage external tools, and iteratively refine objectives through conversation~\cite{huang2025towards}. This enables complex tasks beyond simple top-$n$ recommendations~\cite{xu2025tapping, huang2025towards}. 

Recommender systems are multistakeholder platforms that must balance competing objectives across multiple stakeholders: consumers seeking relevant recommendations, providers seeking visibility, regulators ensuring compliance, and affected communities concerned with broader societal impacts~\cite{burke2017multisided, burke2025centering, ekstrand2025recommending}. Each stakeholder has distinct fairness concerns and success criteria~\cite{burke2025centering, deldjoo2024fairness}. Rather than optimizing for all objectives simultaneously, multi-agent systems can dedicate separate agents to advocate for different stakeholders~\cite{liu2025survey, tran2025multi}. This offers distinct advantages for stakeholder alignment but introduces critical challenges~\cite{liu2025survey, dignum2025agentifying}. Collective decision-making between LLM agents remains opaque and susceptible to failure modes such as confirmation bias, hallucinations, positional bias toward agents who ``speak'' first, and dominance by more articulate agents~\cite{bellina2026conformity, wynn2025talk, chun2025multi, dignum2025agentifying}. Decisions are often distributed, emergent, and context-dependent. This can obscure accountability, cause concerns about transparency and fairness, amplify biases, and create privacy risks~\cite{tran2025multi}.

In this work, we focus on the central challenge of fair and transparent preference aggregation from multiple stakeholder-representing agents. We narrow this gap by adapting social choice theory for collective decision-making to LLM-based multi-agent systems. Social choice methods~\cite{arrow1951social, sen1970collective} have proven theoretical properties to guarantee certain fairness criteria mathematically. 

This paper makes three contributions toward the design of fair multi-agent multistakeholder personalization systems:

\begin{enumerate}
    \item \emph{Related Work and Research Challenges.} In Section~\ref{section2}, we synthesize prior work on multistakeholder fairness and alignment in personalized recommender systems and multi-agent AI systems, preference-aggregation strategies from social choice theory, and stakeholder-centric evaluation methodologies. From this synthesis, we identify and categorize open research challenges specific to multistakeholder personalization with LLM agents.
    \item \emph{Conceptual Framework.} In Section~\ref{section3}, we propose a conceptual framework for fair multi-agent and multistakeholder personalization. The framework combines LLM-based stakeholder alignment with social choice mechanisms (e.g., voting rules) for transparent preference aggregation. This bridges semantic flexibility with theoretical rigor, enabling context-dependent fairness instantiation~\cite{aird2024dynamic, ekstrand2025recommending} while maintaining traceability. Stakeholder-centric evaluation spans individual agent reliability, system-level fairness and accuracy, and stakeholder perception through qualitative studies.    
    \item \emph{Domains and Datasets.} In Section~\ref{section4}, we analyze potential application domains (e.g., tourism, education, healthcare) and identify stakeholder tensions and fairness concerns. For each domain, we align stakeholders to our framework's generic LLM agents. Finally, we outline available datasets suitable for the framework's operationalization and empirical validation. 
    \end{enumerate}

\section{Related Work and Research Challenges}\label{section2}
In this section, we outline current approaches to multistakeholder fairness in multi-agent personalization systems and identify key research challenges that motivate our conceptual framework.


\subsection{Multistakeholder Fairness and Biases in Multi-Agent AI Systems}

A fundamental challenge in aligning AI systems with human values is determining whose conceptions of fairness are represented. Fairness in machine learning and AI tends toward narrow, mathematically tractable formulations~\cite{deldjoo2024fairness, bauer2025conversational, kaya2025mapping}. These fairness formulations are typically rooted in Western ethical theories, aiming to maximize aggregate outcomes, ensure equal opportunity, infer user character, or follow fixed rules~\cite{smith2023many}. Scholars increasingly advocate pluralistic approaches that incorporate non-Western epistemologies, prioritizing collective well-being over individual agency~\cite{mhlambi2023decolonizing, smith2023many}.

Multistakeholder fairness in recommender systems involves consumers, providers, platform-mediated actors (developers, third parties), and upstream/downstream parties~\cite{burke2025centering}. 
Deldjoo~\cite{deldjoo2025understanding} investigates the trade-offs between accuracy and provider-side fairness through prompt design strategies in LLM-based personalization. Fairness-oriented prompts recommend newer items with broader genre distributions, but significantly reduce accuracy. Notably, embedding fairness into system roles (e.g., ``act as a fair recommender'') proves more effective than fairness directives within prompts. These findings underscore the difficulty of reconciling competing objectives within a single agent~\cite{mishra2023ai, liu2025survey, tran2025multi}. 

LLM agents in multi-agent AI systems can act continuously and flexibly in response to context, as opposed to static agents, where agent behavior is defined through explicit logic~\cite{abou2025agentic, dignum2025agentifying}. This can introduce additional complexities. Recent work reveals that LLM agents exhibit conformity and social influence effects that amplify biases~\cite{bellina2026conformity}. Competitive incentives increase hallucination and behavioral misalignment~\cite{el2025moloch}. Interactional fairness (i.e., interpersonal treatment and the honesty and adequacy of explanations) varies depending on the model and whether agents are prompted collaboratively or competitively~\cite{binkyte2025interactional}. Persona-induced biases and in-group favoritism are present across current model generations~\cite{li2026single}. Fairness in multi-agent AI systems emerges from how agents interact, whose values are prioritized, and how conflicts are resolved. This calls for systematic approaches to stakeholder alignment and operationalization in multi-agent AI systems.

\subsection{Stakeholder Alignment and Integration into LLM Agents}
Translating abstract stakeholder values into concrete agent objectives remains a central challenge in multi-agent personalization~\cite{uchoa2025multi}. From a technical perspective, personalization and alignment can be implemented at multiple levels: input (profile-augmented prompting), model (fine-tuning), and objective (aligning to user preferences via Reinforcement Learning from Human Feedback (RLHF) or inference-time methods such as weight merging and model ensembling)~\cite{liu2025survey}. However, RLHF tends to treat disagreements among evaluators as noise, limiting its ability to reflect diverse human values. To address this, Maura et al.~\cite{maura2025jackpot} propose maximal lotteries, a probabilistic social choice rule that better operationalizes human intentions and respects majority preferences.

Recent approaches attempted to encode stakeholder values into agent behavior. Banerjee et al.~\cite{banerjee2025collab} instantiate multiple LLM agents in a tourism recommender by encoding each agent's role, objectives, and ranking rules into system prompts. Uchoa et al.~\cite{uchoa2025multi} propose a privacy-preserving framework. Agents (e.g., students, parents, institutions) generate recommendations based on local policies and use a dedicated negotiation agent to build consensus. Yet, these methods assume that stakeholder values can be cleanly articulated as static policies. The process of eliciting, validating, and refining stakeholder values remains underexplored. Ekstrand et al.~\cite{ekstrand2025recommending} advocate for participatory and community co-design, highlighting that system design is often controlled by developer teams whose values may not align with those of other affected stakeholders. 

\subsection{Multistakeholder Preference Aggregation}
In multi-agent AI systems, agents typically coordinate through natural language, with coordination patterns emerging implicitly from large-scale learning~\cite{dignum2025agentifying, abou2025agentic}. In many cases, a central orchestrator manages several specialized agents, or agents coordinate through negotiation~\cite{abou2025agentic}. While these approaches offer flexibility, they lack verifiable reliability and exacerbate biases and failure modes~\cite{abou2025agentic, wynn2025talk, chun2025multi, bellina2026conformity}. Iterative negotiation approaches have shown promise in balancing competing objectives but face challenges, including resource intensity and convergence on incorrect solutions~\cite{chun2025multi, wynn2025talk}. Banerjee et al.\cite{banerjee2025collab} integrate iterative LLM negotiation with voting-based mediation, finding that the choice of aggregation strategy critically shapes hallucination rates and stability.

To address the limitations of natural language negotiation, we turn to \emph{social choice theory}~\cite{arrow1951social, sen1970collective}. Social choice theory provides theoretically and mathematically grounded methods for aggregating individual preferences into transparent and equitable collective decisions~\cite{dignum2025agentifying}. It is well-established in multi-agent consensus building and resource allocation~\cite{dignum2025agentifying}, and in aggregating preferences for group recommendations~\cite{popescu2013group}. Aird et al.~\cite{aird2024dynamic} operationalize social choice for provider-side fairness in recommendation tasks. In their framework, multiple fairness concerns are assigned to respective agents and dynamically integrated via allocation and aggregation mechanisms. Uchoa et al.~\cite{uchoa2026natural} compare LLM-based mediation with social choice aggregation, showing that mediation can favor marginalized groups, while aggregation tends to optimize overall utility. However, their comparison is limited to a single aggregation mechanism, despite evidence that different voting rules yield vastly different outcomes~\cite{aird2024dynamic}. This underscores the need for systematic exploration of how social choice mechanisms can complement or replace LLM-based mediation to achieve both fairness and explainability.

\subsection{Stakeholder-Centric Evaluation}
Evaluation in multi-agent systems must address individual agents and emerging behaviors across agents. Fairness in AI has traditionally been evaluated in centralized, single-agent settings via offline evaluation~\cite{deldjoo2024fairness}. However, such approaches struggle to capture fairness as contextual and dynamic~\cite{deldjoo2024fairness, mullner2025multistakeholder, aird2024dynamic}. They fail to account for emergent interaction dynamics such as cascading errors, or conflicting objectives that arise in multi-agent settings~\cite{dignum2025agentifying}. 

Grounding mechanisms can be used to evaluate the reliability of individual agents. This can be done by restricting recommendations to verified knowledge bases~\cite{banerjee2025collab}, user validation studies~\cite{gonzalez2025llm}, or similarity-based item matching~\cite{deldjoo2025understanding}. Relevance and diversity metrics can be used to evaluate the performance of aggregate recommendations~\cite{banerjee2025collab, deldjoo2025understanding}. Uchoa et al.~\cite{uchoa2025multi} introduce dedicated coordination agents that resolve conflicts, and audit agents that monitor for bias and policy drift. Aird et al.~\cite{aird2024dynamic} apply the $L_{1/2}$ norm as a metric to balance per-agent fairness with low disparity across agents. Yet, how to effectively integrate agent- and system-level evaluations remains an open challenge.

\subsection{Summary and Research Challenges}
Despite growing interest in multi-agent multistakeholder personalization, existing frameworks do not jointly address the full scope from stakeholder alignment and operationalization to fair aggregation and evaluation in a principled, domain-agnostic manner. On the aggregation side, Uchoa et al.~\cite{uchoa2026natural} offer a first empirical comparison of LLM-based mediation and social choice aggregation. Nonetheless, a broader and more systematic evaluation of social choice voting mechanisms across domains and stakeholder configurations is still lacking. The following, overarching research challenges express these goals and guide our conceptual framework:
\begin{itemize}
    \item \emph{RC1: Aligning abstract stakeholder values with concrete technical objectives of (LLM) agents}
    \item \emph{RC2: Fair aggregation of diverse fairness concerns and consensus building}
    \item \emph{RC3: Stakeholder-centric evaluation of multi-agent personalization systems}
\end{itemize}

\section{Conceptual Framework}\label{section3}

\begin{figure*}[t]
  \centering
        \includegraphics[width=\textwidth]{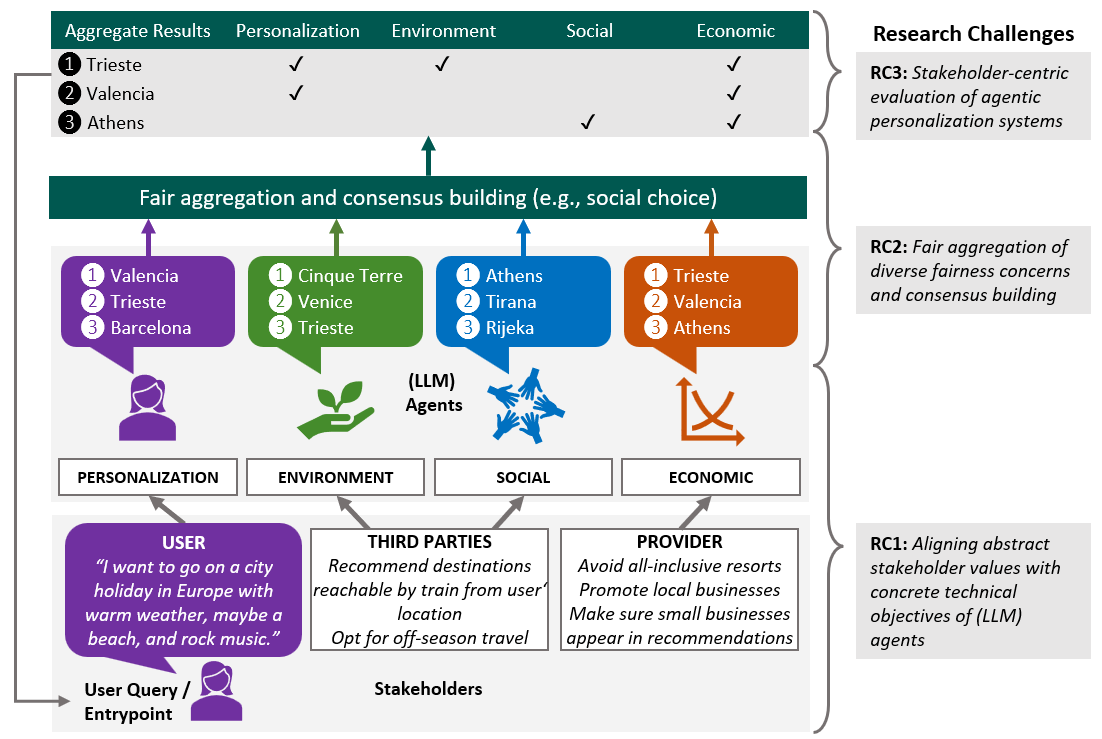}
    \caption{Our conceptual framework for multi-agent multistakeholder personalization in tourism. First, the user enters a query. Other stakeholder values are elicited beforehand. Next, agents are aligned with stakeholders (\emph{RC1}) and each agent generates a candidate set, including justifications for their decision, if applicable. Candidate lists are fed into an aggregation mechanism, and consensus is built, e.g., through different social choice voting mechanisms (\emph{RC2}). Finally, aggregate results and justification of whether different stakeholder objectives are met are returned to the user (\emph{RC3}).}
    \label{framework}
  \vspace{2mm}
\end{figure*}

Next, we represent our conceptual framework for multi-agent multistakeholder personalization systems. Figure~\ref{framework} illustrates the framework using a tourism use case as an example, showcasing the previously identified research challenges. Additional practical considerations are provided in Section~\ref{implementation}. 
Overall, our framework is configurable and domain-agnostic, allowing researchers, practitioners, and communities to instantiate the fairness concerns and aggregation strategies most relevant to their specific context and stakeholder needs. Now, we investigate each research challenge in detail.

\paragraph{RC1: Aligning abstract stakeholder values with concrete technical objectives of (LLM) agents.}
In Figure~\ref{framework}, a user enters queries in natural language, while other relevant stakeholders are identified in advance. Our framework focuses on three core stakeholders: \emph{users}, \emph{providers}, and \emph{third parties}. Each stakeholder is represented by at least one LLM agent, with the flexibility to subdivide stakeholders (e.g., separating third parties into multiple agents) or include additional ones as needed. We exemplify this by mapping stakeholders to the three axes of sustainability (social, economic, environmental)~\cite{lozano2008envisioning}. Stakeholder values are then turned into measurable agent objectives. For example, Forster et al.~\cite{forster2025exploring} frame overtourism mitigation and fairness for niche users as a popularity bias problem, and Aird et al.~\cite{aird2024dynamic} operationalize different provider- and consumer-side fairness concerns into agents, each with its own fairness metrics. The recommendations are then ranked based on each agent's fairness state and relevance.

\paragraph{RC2: Fair aggregation of diverse fairness concerns and consensus building.}
Each agent generates a candidate list based on its objectives and policies, optionally including text-based justifications. We leverage social choice theory to aggregate the agents' candidate lists. Different voting mechanisms are compared, and each stakeholder's influence on the final ranking is quantified. Building on prior work, we employ voting rules including Borda (ordinal scoring), Copeland and Ranked Pairs (pairwise majority), and Kemeny variants (distance minimization)~\cite{aird2024dynamic, uchoa2026natural, lederer2024squared}.
Furthermore, social choice raises important design questions: whether voting occurs simultaneously or sequentially, and whether votes are visible or blind to other agents. These choices can affect strategic behavior~\cite{dignum2025agentifying}. Therefore, our framework keeps an agent's underlying policies and preference model invisible to other agents. 
While social choice methods provide formal guarantees and mathematical traceability, we plan to explore LLM-based mediation as a complementary approach, which may better capture nuanced semantic relationships between agent justifications~\cite{uchoa2026natural}.

\paragraph{RC3: Stakeholder-centric evaluation of multi-agent personalization systems.} 
Evaluation can be performed for individual agents and across multiple agents. At the individual level, we verify agent reliability through grounding mechanisms and monitor for biases, penalizing unreliable agents~\cite{banerjee2025collab, uchoa2025multi}. At the multi-agent level, we assess computational cost and latency alongside recommendation quality metrics, and long-term policy drifts not traceable in single interactions~\cite{uchoa2025multi, bauer2025conversational}. The agents' top-$n$ recommendations can be evaluated collectively and individually to assess contribution and fairness-accuracy trade-offs. Quality metrics include accuracy metrics (e.g., nDCG, recall~\cite{deldjoo2025understanding}) and diversity metrics to account for item fairness (e.g., Gini, normalized entropy~\cite{banerjee2025collab, deldjoo2025understanding}). Different fairness objectives can be defined for each agent~\cite{aird2024dynamic}.
On the user side, popularity bias can be assessed through PopLift~\cite{abdollahpouri2019impact}. 
For the proportional representation of protected groups, Kullback-Leibler or Jensen-Shannon divergence can be used to compare historical and recommended distributions~\cite{lesota2021analyzing}. Furthermore, fairness regret is the difference between an agent's perfect fairness (defined by their objectives~\cite{aird2024dynamic}) and the fairness of a given choice. Beyond offline evaluation, real stakeholder perceptions of conversation quality, recommended content, and behavioral outcomes (satisfaction, willingness to act on recommendations) can be evaluated through user studies and focus groups.

\subsection{Practical Considerations}\label{implementation}
This section explains further details that we omitted from Figure~\ref{framework}, for the sake of simplicity. First, users enter queries in natural language. The personalization agent can refine these through multi-turn dialog. When available, the system incorporates user interaction histories, including past reviews and semantic information about consumed items. The other stakeholder values are identified in advance, either from existing research or through surveys, interviews, or co-design settings. Their values are elicited through personas in natural language~\cite{banerjee2025collab}, opinion surveys~\cite{uchoa2026natural}, or hybrid methods. Based on this, each LLM agent operates with a task-specific system prompt encoding its stakeholder objective. Agents have access to retrieval tools (e.g., vector database search over item embeddings and web search for real-time information such as opening hours or pricing) and a candidate generation module. However, not every agent needs to be LLM-based. In domains where strong conventional recommender systems and rich user profiles already exist, a traditional recommendation model may replace the agent. An agent can create ranked candidate lists without natural language reasoning. A provider-side agent that ensures fair item exposure could leverage rule-based filtering based on statistical parity. Managing the computational cost and latency of running multiple (LLM-based) stakeholder agents remains challenging~\cite{banerjee2025collab}. To balance fairness coverage with computational efficiency, stakeholder agents can be invoked either statically (pre-defined for all queries) or dynamically (activated based on query relevance). Agents are invoked when their compatibility with the current query is high or when the system's fairness performance for that agent's objective falls below a threshold~\cite{aird2024dynamic}. Conversely, if an agent's objectives have been consistently met in recent queries, it may be skipped. This requires maintaining statistics of per-agent fairness metrics across queries. 

\section{Possible Application Domains, Use-Cases, and Available Datasets}\label{section4}

\begin{table*}[!ht]
\small
\caption{Overview of possible application domains with diverse stakeholder settings and fairness concerns. Section~\ref{section4} details how our multi-agent multistakeholder personalization framework addresses these concerns through stakeholder-specific agents and social choice aggregation.}
\label{tab:domains}
\begin{tabularx}{\textwidth}{>{\raggedright\arraybackslash}p{2cm} X X X}
\toprule
\textbf{Domain} & \textbf{Use Case} & \textbf{Stakeholders} & \textbf{Fairness Concerns} \\
\midrule
Tourism~\cite{banerjee2025collab, banerjee2023review, bauer2025conversational} & Destination, point-of-interest, and trip itinerary recommendation & Travelers, local businesses, platform developers, destination managers, communities, ecology & Ensuring personalization and provider visibility, maximizing commissions, mitigating overtourism and environmental degradation \\
\addlinespace
Music~\cite{hadziarapovic2025integrating, unger2025deep} & Balanced streaming recommendations, fair royalty remuneration for all rights holders & Artists, music labels, streaming services, listeners, concert venues & Balancing monetary rewards and market development, ensuring artist exposure, fostering customer loyalty, guaranteeing fair royalties \\
\addlinespace
Education~\cite{uchoa2025multi, burke2025centering, bauer2025conversational} & Course material recommendation & Students, parents, tutors, school management, authorities & Balancing content exposure for tutors, optimizing learning outcomes for students, ensuring curriculum compliance, respecting parental oversight \\
\addlinespace
Digital Archives~\cite{atzenhofer2024value, atzenhofer2025multistakeholder} & Digitization and dissemination of cultural heritage information & Archivists, curators, librarians, platform developers, researchers, students, publishers & Making high-quality information easily accessible, increasing user satisfaction, promoting platform growth, disseminating accurate results, supporting funded curations \\
\addlinespace
Job Market and Human Resources~\cite{kaya2025mapping, langer2023introducing} & Multi-objective job matching, fair algorithmic human resource management & Job seekers, recruiters, companies, public employment services, labor unions, human rights agencies & Supporting personal development and adequate remuneration, promoting economic development, reducing unemployment rates, filling open positions, enabling efficient decision-making \\
\addlinespace
Healthcare~\cite{rozenblit2025toward, thiebes2023design, bauer2025conversational} & Clinical decision support, consumer health, facilitation of stakeholder collaboration & Clinicians, researchers, legal and policy experts, patients, consumer advocates, AI scientists, industry leaders & Ensuring representativeness of training data, maintaining system transparency, disclosing biases, building user trust, empowering patients \\
\addlinespace
Finance~\cite{smith2023many, aird2024dynamic} & Fair matching of loans with lenders & Lenders (platform end users), lending partners (NGOs that mediate international loans), borrowers (individuals and small groups advertising their & Advancing financial inclusion, aligning user interest in low-risk projects, maximizing benefits by investing in low-purchasing-power countries, ensuring fair exposure of borrowers \\
\bottomrule
\end{tabularx}
\end{table*}

This section outlines key domains, potential datasets, and use cases to implement our conceptual framework for multi-agent multistakeholder personalization systems. Table~\ref{tab:domains} identifies stakeholders and their fairness concerns based on existing literature in multistakeholder fairness. Table~\ref{tab:datasets} provides an overview of available datasets used in multistakeholder systems. 

\subsection{Domains and Use Cases}
\paragraph{Tourism.}
Destination, point-of-interest, and trip itinerary recommendations are inherently a multistakeholder problem. Users want to receive personalized recommendations, providers such as hotels and restaurant owners want their businesses to be recommended, local communities and NGOs want to avoid overtourism and environmental degradation, and booking platforms want to receive commissions~\cite{banerjee2023review}. Examples from tourism literature show that stakeholder-specific fairness concerns can be complex and context-based~\cite{mullner2025multistakeholder}. 
A trip itinerary user study~\cite{gonzalez2025llm} with a single-agent, user-focused approach shows that LLM agents can successfully incorporate user histories and generate relevant recommendations. Users, however, report issues with originality (bias toward popular places), time optimization (too few recommendations for the available time), and lack of real-time information. Banerjee et al.~\cite{banerjee2025collab} encode popularity, sustainability, and personalization in separate LLM agents to produce balanced destination recommendation lists. In Section~\ref{section3}, we describe the operationalization of our conceptual framework in tourism in further detail.

\paragraph{Music.}
There are different stakeholder needs in streaming services and music licensing. Copyright policies often treat lyricists, composers, and publishers as a single group, despite their differing needs. Automated licensing systems tend to favor large publishers who can easily navigate them, while small rights holders are left at a disadvantage~\cite{hadziarapovic2025integrating}. For instance, users want to receive relevant song recommendation lists. These can be measured by the ratio of saved or repeated songs. Musicians want fair exposure on the platforms to gain new fans and keep existing listeners. Platforms want to mediate both and generate revenues~\cite{unger2025deep}. Our framework addresses recommendation fairness and licensing equity by instantiating user agents optimizing for relevance, separate artist agents for lyricists, composers, and publishers, ensuring fair exposure and rights distribution, and platform agents balancing engagement with revenue. By encoding distinct preferences for each stakeholder and using social choice aggregation, our framework can counteract the homogenization that disadvantages smaller stakeholders.

\paragraph{Education.}
Value alignment in education entails diverse cultural and philosophical values, competing curricula, and time restrictions~\cite{uchoa2025multi}. In an online educational resource recommendation task, our framework would align the personalization agent with the student's learning preferences; family, privacy, and institutional requirements (schools, authorities, exam schedules) would be separate third-party agents, and fair exposure to learning content from different tutors would be ensured by a provider agent. 

\paragraph{Digital Archives.}
Different stakeholders in digital archives, such as archivists, curators, platform developers, researchers, students, and publishers, have different requirements. Scholarly stakeholders may not be interested in popular or serendipitous recommendations and may favor explicit user control over system-driven personalization~\cite{atzenhofer2025multistakeholder, atzenhofer2024value}. Our framework instantiates separate user agents aligned with scholarly versus discovery-oriented objectives, with additional agents representing curatorial policies (e.g., promoting underexposed collections) and platform goals.

\paragraph{Job Market and Human Resources.}
Job-seeking platforms must balance fairness across multiple stakeholders. For instance, it would be unfair if management positions were predominantly recommended to male candidates, or if job postings from large corporations received disproportionate visibility over start-ups~\cite{kaya2025mapping}. Our framework would deploy user agents for job seekers and provider agents for companies and recruiters. Additional policies from third parties, such as public employment services and labor unions, are also instantiated in separate agents. 
Similarly, algorithmic human resource management systems for task assignment and promotion decisions involve competing stakeholder interests. In our framework, agents for system users (managers and businesses) seek streamlined decision-making and may prefer opacity to prevent gaming. Employee agents want fair exposure to advancement opportunities and transparency in decision-making. Third-party agents represent developer constraints (balancing transparency, accuracy, and maintainability) and regulator requirements (auditability, ethics compliance)~\cite{langer2023introducing}. Candidate list generation by each agent allows for explicit enforcement of fairness metrics. Social choice aggregation then combines these lists, allowing auditors to verify that anti-discrimination policies were actually enforced.

\paragraph{Healthcare.}
The implementation of AI in healthcare entails a complex set of stakeholders, often spanning multiple organizations and countries with different cultural values~\cite{thiebes2023design}. Patients' interests are frequently marginalized in decision-making processes, necessitating robust advocacy in policy formation and system design~\cite{rozenblit2025toward}. Rozenblit et al.~\cite{rozenblit2025toward} advocate for a multistakeholder consortium for effective AI governance in healthcare, including patients, clinicians, ethicists, researchers, and industry leaders, aiming to create patient-centered standards through voting-based consensus mechanisms. If implemented within a clinical decision support system, our framework would include healthcare professionals as users, while patients and advocacy groups are represented by third-party agents. In patient-facing applications, provider agents represent pharmaceutical companies, medical device manufacturers, and healthcare institutions, while third-party stakeholder agents represent ethicists. Unlike informal consortium voting, our social choice mechanisms provide formal guarantees about how stakeholder preferences combine, addressing the accountability gaps between governmental bodies and private corporations~\cite{rozenblit2025toward}.

\paragraph{Finance.}
Peer-to-peer microlending platforms like Kiva.org involve multiple stakeholders with different objectives~\cite{smith2023many, aird2024dynamic}. Lenders seek low-risk or high-impact loans, while borrowers need equitable exposure. The platform must balance both while proving accountability to funding institutions~\cite{smith2023many}. Our framework represents lenders as users with configurable risk/impact preferences, borrowers through provider-side agents enforcing exposure equity (e.g., geographic or sector diversity), and policies of platform developers and intermediaries through third-party agents. Social choice aggregation allows the provider and third-party agents to ensure minimum exposure thresholds for underrepresented borrower categories.
\clearpage

\subsection{Datasets}
Table~\ref{tab:datasets} provides an overview of relevant datasets for multistakeholder settings. Personalized alignment of LLMs requires high-quality datasets containing both implicit and explicit user-item interactions~\cite{liu2025survey, lin2025can}. Our framework particularly leverages rating-based datasets with contextual item information where LLMs can process semantic content (e.g., categorical attributes, natural language descriptions) that traditional rating-based models cannot effectively utilize. Such datasets typically contain user profiles and relationships (demographics, social networks), historical dialogues with LLM agents, user-generated content (reviews), interaction history (ratings, consumed items), and pre-defined user preferences~\cite{liu2025survey}. Dataset selection often lacks rigorous justification. Vente et al.~\cite{vente2025aps} address this gap by introducing a web tool that analyzes the similarity of 96 recommender system datasets to inform selection decisions, noting that most experiments rely on only four datasets without a clear rationale. The scarcity of high-quality, domain-specific datasets with sufficient depth and questions about training data leakage~\cite{di2025llms} have driven researchers toward synthetic data generation. For instance, Banerjee et al.~\cite{banerjee2025synthtrips} introduce a framework that produces diverse travel queries from persona-based preferences and sustainability filters, grounded in an external knowledge base. This approach provides a blueprint for combining real-world and synthetic datasets to address multistakeholder fairness in multi-agent AI systems across domains. 

\begin{table*}[]
\small
\begin{threeparttable}
\caption{Overview of potential datasets for future experiments with descriptions and references to prior work. Columns indicate whether datasets are synthetic, contain user-item ratings, or include rich contextual information.}
\label{tab:datasets}
\begin{tabularx}{\textwidth}{>{\setlength{\parindent}{0pt}\arraybackslash}m{1.8cm} >{\setlength{\parindent}{0pt}\arraybackslash}m{6.6cm} >{\centering\arraybackslash}m{1.2cm} >{\centering\arraybackslash}m{1.3cm} >{\centering\arraybackslash}m{1.2cm} >{\centering\arraybackslash}m{1.3cm}}
\toprule
\textbf{Dataset} & \textbf{Description} & \textbf{Domain} & \textbf{Synthetic} & \textbf{Ratings} & \textbf{Item Context} \\
\midrule
SynthTRIPs\tnote{1} & Repository with synthetic user queries for personalized travel planning and knowledge base of European cities & Tourism \cite{banerjee2025collab} & \checkmark &  & \checkmark \\
\addlinespace
Microlending\tnote{2} & Lender-borrower interactions and categorical loan features & Finance \cite{aird2024dynamic} &  & \checkmark & \checkmark\\
\addlinespace
Context Trails\tnote{4} & Dataset and code to enrich Foursquare user-item interaction datasets with additional contextual information (opening times, fine-grained venue information, weather data) & Tourism \cite{sanchez2025context}  &  & \checkmark & \checkmark\\
\addlinespace
Educational Stakeholder Archetypes\tnote{5} &  Stakeholder archetypes with preferences on a five-point scale generated from stakeholder persona-based natural language prompts & Education \cite{uchoa2026natural} & \checkmark & &  \\
\addlinespace
Nemotron Personas\tnote{6}& Synthetic dataset featuring personas from different countries, including their professional, sports, cultural personas & Various & \checkmark &  & \\
\addlinespace
Yelp\tnote{7} & User-item interactions for points-of-interest, ratings and semantic reviews, additional item metadata (e.g., business category, opening times)  & Tourism \cite{rahmani2022unfairness} &  & \checkmark &  \checkmark \\
\bottomrule
\end{tabularx}
\begin{tablenotes}
\footnotesize
\item[1] \url{https://ashmibanerjee.github.io/synthTRIPS-website/}
\item[2] \url{https://scholar.colorado.edu/concern/datasets/j6731518r}
\item[3] \url{https://grouplens.org/datasets/movielens/}
\item[4] \url{https://zenodo.org/records/15855966}
\item[5] Data available upon request from the authors~\cite{uchoa2026natural}
\item[6] \url{https://huggingface.co/datasets/nvidia/Nemotron-Personas-Singapore}; Brazil, India, USA, France, and Japan also available
\item[7] \url{https://business.yelp.com/data/resources/open-dataset/}
\end{tablenotes}
\end{threeparttable}
\end{table*}

\section{Conclusion and Future Work}
In this paper, we identify three core research challenges for multi-agent multistakeholder personalization systems: aligning abstract stakeholder values with concrete objectives of LLM agents, aggregating diverse fairness concerns, and conducting stakeholder-centric evaluation. We propose a conceptual framework that addresses these challenges to enable multistakeholder and multi-agent fairness in personalization systems. Leveraging social choice theory for aggregation ensures transparent and accountable consensus-building across diverse deployment domains. We illustrate our framework with a tourism recommendation use case. Additionally, we provide examples of domain-specific use cases, stakeholders, fairness concerns, and datasets, and discuss the applicability of our framework for various domains.

In the future, we plan to validate our conceptual framework empirically. This includes assessing how stakeholder values can be effectively integrated and evaluated. Moreover, we aim to compare various social choice mechanisms with LLM-based mediation and explore hybrid aggregation approaches that combine both paradigms. We, for instance, envision mixed strategies where social choice mechanisms ensure formal fairness guarantees and mathematical traceability of stakeholder influence. At the same time, LLMs provide natural language justifications that explain which stakeholder concerns were prioritized and what trade-offs were made. This approach could address limitations of pure social choice (lack of semantic explanations) and pure LLM mediation (lack of formal guarantees). Additionally, we plan to compare our framework against a non-LLM baseline (e.g., aggregating traditional recommendation lists with social choice) to quantify the added value of LLM multi-agent personalization.

\begin{acknowledgments}
This work is conducted within the Interfaces of Agent-Centric Artificial Intelligence (IACAI) COMET module, funded by the Austrian Research Promotion Agency (FFG). 
\end{acknowledgments}

\section*{Declaration on Generative AI}
During the preparation of this work, the authors used Grammarly for grammar checks, spelling checks, and sentence polishing. After using this tool, the authors reviewed and edited the content as needed and take full responsibility for the publication’s content.

\bibliography{main.bib}

\end{document}